\documentclass[twocolumn,amsmath,amssymb,prl]{revtex4-1}                    %Reference without titles
\usepackage{titlesec}
\usepackage[colorlinks,bookmarks=false,citecolor=blue,linkcolor=red,urlcolor=blue]{hyperref}
\usepackage{epsfig}
\usepackage{color}
\usepackage{subfigure}
\usepackage{graphicx}    % Include figure files
\usepackage{dcolumn}     % Align table columns on decimal point
\usepackage{lipsum}      % to move around 2-column tables !
\input{insbox}

\newcommand{\be}{\begin{equation}}
\newcommand{\ee}{\end{equation}}

\definecolor{drkgr}{rgb}{0.05,0.6,0.2}

\begin{document}

\title{Yb$^{3+}$ $f$-$f$ excitations in NaYbSe$_2$: benchmarking embedded-cluster quantum chemical schemes for 4$f$ insulators}

\author{P.~Bhattacharyya}
\affiliation{Institute for Theoretical Solid State Physics, Leibniz IFW Dresden, Helmholtzstrasse~20, 01069 Dresden, Germany}

\author{L.~Hozoi}
\affiliation{Institute for Theoretical Solid State Physics, Leibniz IFW Dresden, Helmholtzstrasse~20, 01069 Dresden, Germany}

\begin{abstract}
\noindent
$\tilde{S}\!=\!1/2$ triangular-lattice $f$-electron materials define a dynamic research area in
condensed matter magnetism. 
In various Yb 4$f^{13}$ triangular-lattice compounds, for example, spin-liquid ground states seem to
be realized.
Using {\it ab initio} quantum chemical methods, we here investigate how correlation effects involving
the 4$f$ electrons affect the on-site $f$-$f$ excitation spectrum in NaYbSe$_2$.
The system is well suited for such a study since unambiguous inelastic neutron scattering data are
available for the Yb$^{3+}$ $f$-$f$ transitions.
The excitation energies obtained by configuration-interaction calculations with single and double
substitutions agree within 3-4 meV with experimental values, which provides a not so expensive
alternative to fitting experimental data at the model-Hamiltonian level in order to analyze $f$-center
multiplet structures.
\end{abstract}

\date\today
\maketitle

{\it Introduction.\,}
Triangular-lattice rare-earth ($R$) delafossites with the general formula $ARX_2$ provide a appealing
platform for investigating frustrated quantum magnetism, in particular, the possbility of realizing
quantum spin-liquid ground states \cite{yb112_liu_2018,kitaev_d_f_sasha_2020}.
Aside from geometrical frustration, the strong 4$f$-shell spin-orbit coupling (SOC) and edge-sharing
arrangement of adjacent $RX_6$ octahedra leave room for large intersite exchange anisotropy
% , bond dependence of the anisotropic intersite couplings,
and additional frustration \cite{kitaev_d_f_motome_2020,kitaev_d_f_sasha_2020}.
From an electronic-structure point of view, the simplest 4$f$-shell electron configurations are in
principle 4$f^1$ and 4$f^{13}$, with either one electron or one hole within the set of seven 4$f$
orbitals.
Such 4$f$-shell occupations are realized in, e.\,g., KCe$X_2$ and NaYb$X_2$ compounds, where $X$ can
be O, S, or Se.
Since $S\!=\!1/2$ and $L\!=\!2$, SOC yields $J\!=\!7/2$ and $J\!=\!5/2$ terms, split up in solids due
to crystal-field (CF) effects.
The $J\!=\!5/2$-like states describe the low-energy multiplet structure of Ce$^{3+}$ ions in KCe$X_2$
delafossites while the relevant low-energy states for Yb$^{3+}$ are related to the free-ion $J\!=\!7/2$
term.
In lower symmetries (e.\,g., $D_{3d}$ 4$f$-site symmetry in $ARX_2$ delafossites), the $J\!=\!5/2$-like
spectrum amounts to three Kramers doublets, i.\,e., two low-energy on-site excitations.
Interestingly, three high-intensity peaks are found experimentally in KCeS$_2$ \cite{KCeS2_gael_2020}, KCeO$_2$ \cite{Bordelon_KCeO2}, and RbCe$X_2$ ($X$ = O, S, Se, and Te)
\cite{RbCeX2}, which indicates additional degrees of freedom in the Ce-based
delafossites, presumably strong vibrational couplings.
On the other hand, one excitation is missed in the inelastic neutron scattering (INS) measurements on
NaYbS$_2$ \cite{yb112_baenitz_2018}.
While {\it ab initio} quantum chemical methods were in those cases helpful for achieving a more complete
picture on the underlying multiplet structures \cite{KCeS2_gael_2020,yb112_baenitz_2018}, in order to
test in greater detail the abilities of quantum chemical computational schemes, it is desirable to use
as reference safer sets of experimental data. 
This seems to be achieved in NaYbSe$_2$, for which convergent INS results are reported for the expected
three low-energy excited-state 4$f^{13}$ terms \cite{Zhang_INS,Dai_INS}. 
Here we address the 4$f^{13}$ electronic structure of NaYbSe$_2$ by quantum chemical methods and evidence 
that correlation effects are important.
In particular, it is shown that a rather advanced type of correlation treatment is required in order
to achieve reasonable agreement with experiment for the $f$-$f$ excitation energies.
The analysis provides useful reference computational data in the context of correlation effects in
4$f$-electron insulators, much less investigated by {\it ab initio} quantum chemical techniques in
comparison to  $d$-electron compounds.

%%%%%%%%%%%%%%%%%
%%% FIGURE 1 %%%%
%%%%%%%%%%%%%%%%%
\begin{figure}[b]
\includegraphics[width=1\columnwidth]{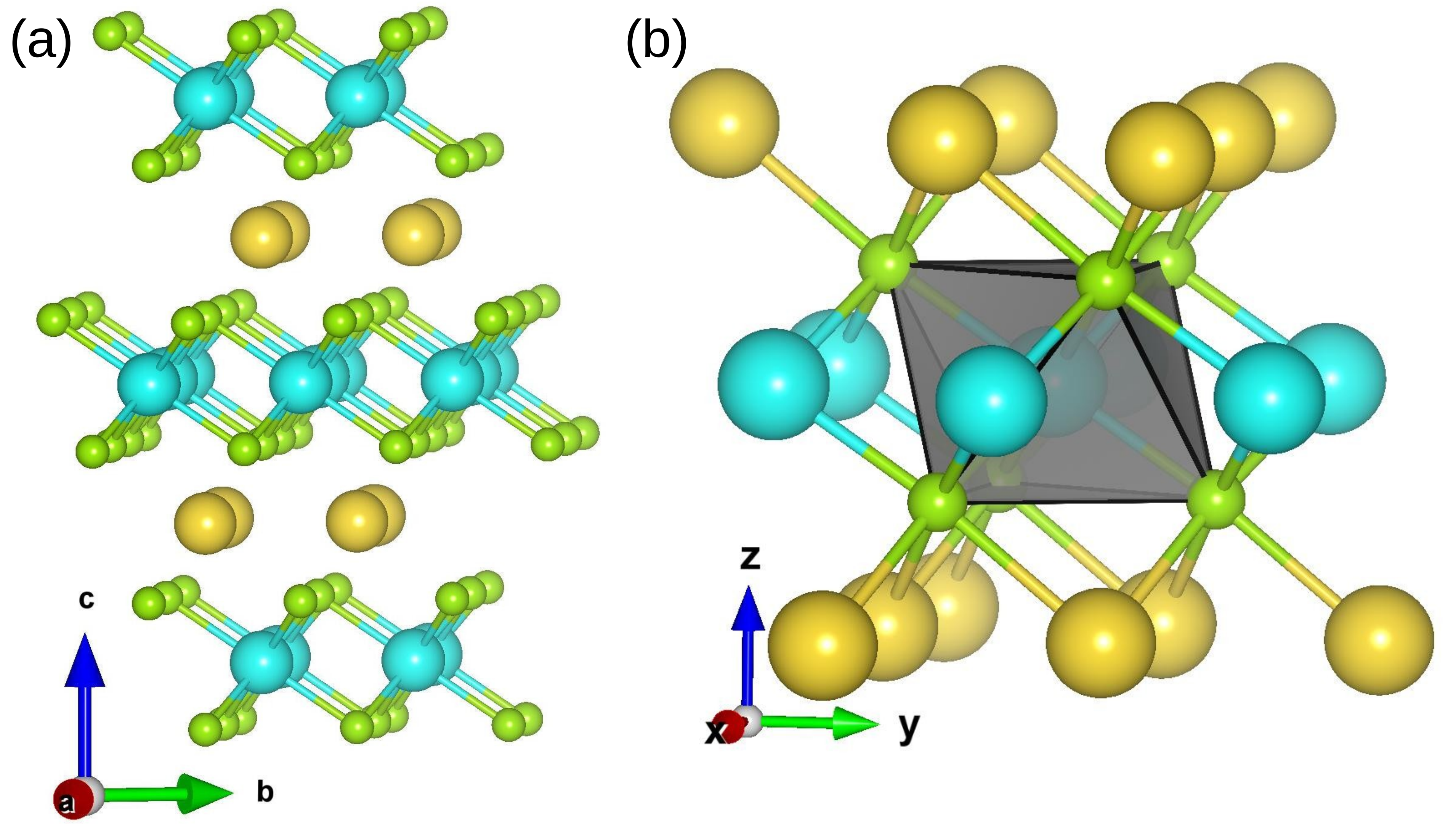}
\caption{
(a) Crystal structure of NaYbSe$_2$ with successive ionic layers, plot using the {\sc vesta} visualization program \cite{vesta}.
(b) Finite cluster considered in the computations.
Cyan, yellow, and green spheres denote Yb, Na, and Se sites, respectively.
}
\label{fig_1}
\end{figure}

%%%%%%%%%%%%%
%% TABLE 1 %%
%%%%%%%%%%%%%
\begin{table*}[!ht!]
\caption{
Yb$^{3+}$ 4$f^{13}$ multiplet structure in NaYbSe$_2$, relative energies in meV.
Notations corresponding to $D_{3d}$ point-group symmetry are employed for the crystal-field (SOC not considered) and spin-orbit
states (+SOC). For the double group,
notations as in Ref. \cite{sugano_tanabe} (e.g., Appendix I in \cite{sugano_tanabe}) are used. Davidson corrections \cite{olsen_bible} were added to the MRCI energies.
}
\begin{tabular}{l c c| c c c l}
\hline
\hline

Crystal-field states &CASSCF  &MRCI   &CASSCF+SOC  &MRCI+SOC   &INS\cite{Zhang_INS}            &Spin-orbit states\\

\hline

$^2\!A_{2u}$         &0       &0      &0           &0          &0              &$\Gamma_6$\\
$^2\!E_u$            &8       &10     &8.0         &12.3       &15.8           &$\Gamma_6$\\
                     &8       &10     &15.8        &20.7       &24.3           &$\Gamma_4+\Gamma_5$\\
$^2\!A_{1u}$         &18      &28     &25.4        &34.6       &30.5           &$\Gamma_6$\\
$^2\!E_u$            &28      &37     &1275.6      &1301.3     &--             &$\Gamma_6$\\
                     &28      &37     &1280.5      &1306.6     &--             &$\Gamma_4+\Gamma_5$\\
$^2\!A_{2u}$         &34      &47     &1296.2      &1328.5     &--             &$\Gamma_6$\\
\hline
\hline
\end{tabular}
\label{tab:excited_energy}
\end{table*}

%%%%%%%%%%%%%
%% TABLE 2 %%
%%%%%%%%%%%%%
\begin{table}[b]
\caption{ 
Ground-state $g$ factors in NaYbSe$_2$ as obtained by CASSCF+SOC, MRCI+SOC, and electron spin resonance (ESR) measurements.
}
\begin{tabular}{l c c c c c l}
\hline
\hline

              &CASSCF+SOC  &MRCI+SOC  &ESR     \\

\hline

$g_{ab}$      &3.25        &3.19      &3.13 \cite{NaYbSe2_Ranjith}, 3.10 \cite{Zhang_ESR}    \\
$g_c$         &0.21        &0.76      &1.01 \cite{NaYbSe2_Ranjith}, 0.96 \cite{Zhang_ESR}    \\
%CASSCF+SOC                    &3.25      &0.21     \\
%MRCI+SOC                      &3.19      &0.76     \\
%ESR \cite{NaYbSe2_Ranjith}    &3.13      &1.01     \\
%ESR \cite{Zhang_ESR}          &3.10      &0.96     \\
\hline
\hline
\end{tabular}
\label{tab:g_factors}
\end{table}

{\it Computational details.\,} 
%The recent finding of layered triangular-lattice 
%ternary lanthanide system YbMgGaO$_4$ \cite{YbMgGaO4_PRX1,YbMgGaO4_PRX2,YbMgGaO4_PRL1,YbMgGaO4_PRL2,
%YbMgGaO4_PRL3,YbMgGaO4_PRL4,YbMgGaO4_Nat_Phys,YbMgGaO4_Nat} as a potential spin-liquid contender has demanded 
%further investigation to search for Yb-based compounds having similar characteristics. Also the trivalent 
%delafossite type Yb-based systems of the family NaYbX$_2$, where X = O, S, Se attracted much attention as it 
%exhibits model quantum disordered states \cite{yb112_bordelon_2019,yb13_ziba_2019,NaYbX2_Liu,NaYbX2_Bordelon,
%yb112_baenitz_2019,NaYbX2_Ding,NaYbX2_Ranjith,NaYbX2_Sarkar,NaYbX2_Sichelschmidt,yb112_baenitz_2018}. 
%
Trigonally distorted, edge-sharing YbSe$_6$ octahedra decorate a two-dimensional triangular magnetic lattice 
in NaYbSe$_2$, as presented in Fig.~\ref{fig_1}(a).
To achieve a correct picture on the Yb$^{3+}$ 4$f^{13}$ multiplet structure in this crystalline environment, 
quantum chemical calculations were carried out using the {\sc molpro} \cite{Molpro} computer program; 
in these computations, a finite cluster composed of a YbSe$_6$ octahedron, six adjacent Yb sites, and twelve Na 
nearest neighbors was considered, as shown in Fig.~\ref{fig_1}(b). This finite fragment was embedded within a 
large array of point charges, which replicate the crystalline Madelung field \footnote{a fully ionic picture 
was assumed, with formal charges of $+1$, $+3$, and $-2$ at the Na, Yb, and Se sites, respectively.}
within the cluster region; this distribution of point charges 
was generated using the {\sc ewald} package \cite{Klintenberg_et_al,Derenzo_et_al}. 
We employed a similar material model with simple point-charge embedding \cite{2203.06394} to study the Ce$^{3+}$ 4$f^1$ multiplet 
structure in NaCeO$_2$ and found excellent agreement between experimental results and our computed $f$-$f$ excitation 
energies and $g$ factors; this convincing confirmation of the material model motivated us to apply a similar methodology for 
investigating the Yb$^{3+}$ 4$f^{13}$ multiplet structure in the NaYbSe$_2$ compound. More sophisticated embedding schemes are based on e.g. density functional theory \cite{embedding_DFT1,embedding_DFT2,embedding_DFT3} or prior Hartree-Fock computations for the periodic system \cite{embedding_HF1,embedding_HF2}.

The quantum chemical study was initiated as complete active space self-consistent field (CASSCF)
calculations \cite{olsen_bible,MCSCF_Molpro}. For this purpose, an active orbital space containing the seven 4$f$ orbitals of the 
central Yb atom was considered. We obtained the seven crystal-field states related to the 4$f^{13}$ manifold using a 
state-averaged variational optimization \cite{olsen_bible}. Next, we performed electron-correlated computations at the level of 
multireference configuration-interaction (MRCI) with single and double excitations (MRSDCI)
\cite{olsen_bible,MRCI_Molpro} out of the Yb 4$f$ and Se 4$p$ orbitals of the central YbSe$_6$ octahedron. At last, 
we carried out spin-orbit calculations following the procedure described in Ref.~\cite{Berning_et_al}. 
The diagonal elements of the spin-orbit matrix in the final MRCI+SOC step were 
replaced by Davidson-corrected \cite{olsen_bible} MRCI energies to
obtain the results enlisted in the fifth column of Table\;\ref{tab:excited_energy}. 
In these computations, for the central Yb ion quasirelativistic pseudopotentials \cite{Dolg_Stoll_Preuss} and 
valence basis sets of quadruple-$\zeta$ quality \cite{Cao_Dolg_1,Cao_Dolg_2} were employed, whereas we used 
all-electron triple-$\zeta$ basis sets for the Se ligands of the central YbSe$_6$ octahedron \cite{Dunning_Se}. 
Large-core pseudopotentials were adopted for the six Yb nearest neighbors \cite{Dolg1989} and also for the twelve 
adjacent Na cations \cite{Fuentealba_Na}. For the former, the 4$f$ electrons are also part of the effective potential.

Crystallographic data were used as provided in Ref.~\cite{Gray_et_al}.
NaYbSe$_2$ exhibits a $R\bar{3}m$ layered structure (space-group number 166) \cite{KCeO2_qc_2020}, see Fig.\;\ref{fig_1}; 
the Wyckoff positions of Na, Yb, and Se are 3a (0,\,0,\,0), 3b (0,\,0,\,1/2), and 6c (0,\,0,\,0.2424), 
respectively, whereas the experimentally determined lattice constants are $a$=$b$=4.0568 \AA \ and $c$=20.7720 
\AA \ \cite{Gray_et_al}. A given YbSe$_6$ octahedron features $D_{3d}$ point-group symmetry, with Yb-Se bond lengths 
of 2.82 \AA ~and as a result of trigonal compression Se-Yb-Se bond angles of 88.07 and 91.93 degrees.

{\it Results and discussion.\,}
Results obtained from CASSCF and MRCI computations without and with spin-orbit coupling are provided in
Table~\ref{tab:excited_energy}, along with INS data.
Without SOC, one $A_{1u}$, two nondegenerate $A_{2u}$, and two sets of doubly degenerate $E_u$ sublevels
are expected in $D_{3d}$ symmetry \cite{atkins_70}, out of which the $^2\!A_{2u}$ crystal-field state
is found to be the lowest in our calculations.
Though SOC defines a dominant energy scale, the sequence of the above mentioned crystal-field levels
and the magnitude of the crystal-field splittings determine the precise nature of the spin-orbit ground
state. 
Additionally, the details of the multiplet structure depend on intra-atomic electronic correlations
and to a smaller extent on $R$\,$f$\,--\,$X$\,$p$ correlation effects.

The computational data summarized in Table\;\ref{tab:excited_energy} show that the MRCI treatment gives
rise to significant corrections, of up to 35\%, on top of the CASSCF approximation.
A post-CASSCF correlation treatment is therefore worth for late rare-earth ions, an aspect previously pointed out in e.g. Refs. \cite{Ungur_chem,Ungur_prb}. 
%
%{\color{red}The energy difference between the low-lying $J=7/2$ ground-state multiplet
%and the excited $J=5/2$ states is as large as $\sim$1.3 eV, while the splittings
%determined by the ionic crystalline environment are in the range of
%only tens of meV.
%From Table \ref{tab:excited_energy}, it is noted that the extent of correlations 
%reflected in MRCI results is comparatively higher in the spin-orbit states as 
%compared to crystal-field states.}
%
The low-lying excitation energies obtained by spin-orbit MRCI are 12.3, 20.7, and 34.6 meV.
Experimentally, on-site $f$-$f$ transitions are observed at 15.8, 24.3, and 30.5 meV in the INS spectrum
\cite{Zhang_INS}.
Comparing the two sets of relative energies, we see that the MRCI+SOC results reproduce the experimental
values within 10-20\%.
%
% Very good agreement with experimental data is observed in the quantum chemical calculations, as also found
% in the case of $d$-electron compounds with one hole per site \cite{Murugesan_et_al,Huang_et_al}.
MRCI results of similar quality were obtained for the on-site multiplet structures in $d$-electron compounds
with one single hole within the valence shell, i.\,e., cuprates \cite{Murugesan_et_al,Huang_et_al}.
It turns out that spin-orbit interactions may be sizable also in those materials \cite{Murugesan_et_al}.

Not only NaYbSe$_2$ \cite{,NaYbSe2_Ranjith,2112.07199}, but also 
NaYbO$_2$ \cite{yb112_bordelon_2019,yb112_baenitz_2019}, NaYbS$_2$
\cite{yb112_baenitz_2018}, KYbSe$_2$ \cite{Xing_et_al}, and RbYbSe$_2$ \cite{Xing_et_al} seem to realize
spin-liquid ground states as no long-range order is observed at sub-Kelvin temperatures in any of
these magnetic systems.
INS data are not available for KYbSe$_2$ and RbYbSe$_2$, while in NaYbS$_2$ one $f$-$f$ excitation is for 
unclear reasons missing in existent INS spectra \cite{yb112_baenitz_2018}.  
%
% We note that the lowest three $f$-$f$ excitation energies in NaYbSe$_2$ are reduced 
Within the NaYb$X_2$ series, the lowest three $f$-$f$ excitation energies are reduced in NaYbSe$_2$
as compared to NaYbO$_2$ and NaYbS$_2$ \cite{yb13_ziba_2019}. 
This feature can be understood on the basis of the larger ligand ionic radius and increased Yb-ligand bond
lengths in NaYbSe$_2$.
Longer Yb-ligand bonds result in smaller ligand-field effects, i.\,e., weaker splittings within the
$J\!=\!7/2$ manifold.
Other crystallographic/electrostatic features affecting the 4$f$-shell multiplet structure are
the trigonal compression of the Yb$X_6$ octahedra and anisotropies related to the layered lattice
configuration.
How such effects may compete with each other was recently analyzed in, e.\,g., Refs.~\cite{yb13_ziba_2019,
KCeO2_qc_2020}.
%
%We figure out that only the lowest computed $f$-$f$ excitation energy in NaYbO$_2$ \cite{yb13_ziba_2019} 
%is in reasonable agreement with INS measurement \cite{yb112_tsirlin_2019}; while for the next two computed 
%transitions, significant differences were observed as compared to the experimental values. We also found that 
%the lowest computed $f$-$f$ transition in NaYbS$_2$ \cite{yb13_ziba_2019} is completely missing in the 
%INS measurement \cite{yb112_baenitz_2018}; however, the lowest two INS peaks corresponding to the 
%free-ion $J=7/2$ term might be very close enough and we were not able to predict them correctly using 
%our earlier quantum chemical study \cite{yb13_ziba_2019}. Although, in this work, all the experimentally 
%measured INS peak positions in NaYbSe$_2$ \cite{Zhang_INS} are probed and also in fair agreement with the 
%experimental values.
%
% The computed crystal-field splittings are nevertheless significantly smaller as compared 
% to those found in KCeO$_2$ \cite{KCeO2_qc_2020,Bordelon_KCeO2}. 

Yb-ion $g$ factors computed on the basis of the spin-orbit CASSCF and MRCI wave functions are displayed in 
Table\;\ref{tab:g_factors}, along with $g$ factors derived by electron spin resonance (ESR) 
measurements \cite{NaYbSe2_Ranjith,Zhang_ESR} and fitting INS peak positions \cite{Zhang_INS}.
It is seen that the $g$ factors are strongly anisotropic, similar to the $g$ factors obtained from ESR
measurements on CsYbSe$_2$ \cite{2106.12451}, NaYbS$_2$ \cite{yb112_baenitz_2018}, and
NaYbO$_2$ \cite{yb112_bordelon_2019,yb112_baenitz_2019}.
% {\color{red}From Table \ref{tab:g_factors}, the effect of correlation 
% is clearly observed, as the MRCI $g$ factors are more closer to the experimental values, especially in case of 
% the perpendicular component of the $g$ factors ($g_c$). Though, the splittings obtained by CASSCF and MRCI are 
% marginally differed, the computed wave functions at MRCI level offered an important role to estimate 
% the $g$ factors.}
The MRCI treatment yields a very large correction to the perpendicular component $g_c$, in particular,
and provides a value that is much closer to the experimental result.
%
% The $g$ factors obtained from ESR measurements on CsYbSe$_2$ \cite{2106.12451} are similar to those found in
% NaYbSe$_2$.
% More recently, significant vibronic effects \cite{CsYbSe2_lawrie_2022} were 
% evidenced in CsYbSe$_2$. Strong vibronic effects were also observed on $g$ factors in the $\Gamma_8$ cubic 
% multiplet \cite{Iwahara_et_al} and CeF$_3$ \cite{Gerlinger_et_al}.
% These can in principle account for the differences observed between the 
% ESR and MRCI $g$ factors.
%
Remaining differences between computational and experimental results may be related to correlation effects not
accounted for at the MRSDCI level but also to vibronic couplings \cite{Iwahara_et_al,Gerlinger_et_al,Thalmeier_et_al}.
Important vibronic effects were recently evidenced in the Raman spectra of both NaYbSe$_2$ \cite{2203.13361} and 
CsYbSe$_2$ \cite{CsYbSe2_lawrie_2022}.
% {\color{red}
% The g factors for other Yb$^{3+}$ delafossites, e.g., NaYbO$_2$ \cite{yb112_baenitz_2019,yb112_bordelon_2019} 
% and NaYbS$_2$ \cite{yb112_baenitz_2018,yb112_bordelon_2019} manifest similar trends of strong anisotropy. 
% }

{\it Conclusions, outlook.\,}
In sum, the performance of a embedded-cluster MRSDCI computational scheme is documented for the case of
a Yb$^{3+}$ 4$f^{13}$ compound, NaYbSe$_2$.
The material is part of the $A$Yb$X_2$ delafossite family, of substantial interest in nowadays research as
platform for quantum spin-liquid states.
It represents a good testing ground for electronic-structure methods because complete, convergent INS
data are available for the low-energy on-site $f$-$f$ transitions \cite{Zhang_INS,Dai_INS}.
While the reliability of the embedding scheme itself was recently proven on Ce$^{3+}$ 4$f^1$ compounds such
as KCeO$_2$ \cite{KCeO2_qc_2020,Bordelon_KCeO2} and NaCeO$_2$ \cite{2203.06394}, we focus in the present
investigation on the extent of electronic correlations on a $RX_6$ octahedron.
While minor for the Ce$^{3+}$ 4$f^1$ octahedral system \cite{KCeO2_qc_2020,2203.06394}, with `minimal'
4$f$-shell occupation, it is shown that correlation effects are important for the Yb$^{3+}$ 4$f^{13}$
valence configuration:
% {\color{red}
% The MRCI treatment brings remarkable corrections, 35\%, 24\%, and 27\%, respectively in addition to 
% the CASSCF excitation energies.
% The computed excitation energies are in good agreement within $\sim$4 meV with the INS measurements.
% }
the MRSDCI treatment brings corrections of up to 35\% to the CASSCF on-site $f$-$f$ excitation energies,
delivering results within 3-4 meV of experimentally determined peak positions. 
A sizable correlation-induced correction is also found for one of the ground-state $g$ factors. 
Extensions toward the computation of full INS spectra (i.\,e., not only excitation energies but also
intensities) for single-site scattering will provide a very effective {\it ab initio} tool for the
analysis and interpretation of $f$-ion multiplet structures in lanthanide/actinide magnetic insulators. 

 \ 

{\it Acknowledgments.\,}
We thank T.~Petersen, M.~S.~Eldeeb, H. Stoll, and U. K. R\"o\ss{}ler for discussions, U.~Nitzsche for technical assistance, and
the German Research Foundation (Project No. 441216021) for financial support.

\bibliography{ff_naybse2_may13}

\end{document}